\documentclass{aa}
\usepackage{hyperref}
\usepackage{graphicx}
\usepackage{natbib  }
\usepackage{color   }
\usepackage{array   }
\usepackage{txfonts }
\usepackage{breakurl}
\usepackage{multirow}
\usepackage{multicol}
\usepackage{booktabs}
\usepackage{dcolumn }

 \def\Jyb{\textrm{Jy~beam}$^{-1}$}        
 \newcommand{\dms}[3]{$#1^{#2}\!\!.#3$}   

 \def\Fermi{{\slshape Fermi}-LAT}         

\begin{document} 

\title{G51.04+0.07 and its environment: Identification of a new Galactic supernova remnant at low radio frequencies}
\titlerunning{G51.04+0.07 a newly recognised Galactic SNR}

\author{L. Supan\inst{1,2}
\and G. Castelletti\inst{1,2}
\and W.~M. Peters\inst{3}
\and N.~E. Kassim\inst{3}}

\offprints{L.  Supan}
\institute{CONICET-Universidad de Buenos Aires, Instituto de Astronomía y Física del Espacio (IAFE), Buenos Aires, Argentina
\and  Universidad de Buenos Aires, Facultad de Ciencias Exactas y Naturales. Buenos Aires, Argentina \\
\email{lsupan@iafe.uba.ar}
\and Remote Sensing Division, Naval Research Laboratory, Washington DC 20375, USA}
\authorrunning{Supan et al.}

\date{Received 9 March 2018 / Accepted 26 April 2018}

\abstract
{We have identified a new supernova remnant (SNR), G51.04+0.07, using observations at 74~MHz from the Very Large Array Low-Frequency Sky Survey Redux. Earlier, higher frequency radio continuum, recombination line, and infrared data had correctly inferred the presence of nonthermal radio emission within a larger, complex environment including ionised nebulae and active star formation. However, our observations have allowed us to redefine at least one SNR as a relatively small source (\dms{7}{\prime}{5}$\times$3$^{\prime}$ in size) located at the southern periphery of the  originally defined SNR candidate G51.21+0.11. 
The integrated flux density of G51.04+0.07 at 74~MHz is 6.1$\pm$0.8~Jy, while its radio continuum spectrum has a slope $\alpha$=$-$0.52$\pm$0.05 ($S_{\nu} \propto \nu^{\alpha}$), typical of a shell-type remnant. We also measured spatial variations in the spectral index between 74 and 1400~MHz across the source, ranging from a steeper spectrum ($\alpha$=$-$0.50$\pm$0.04) coincident with the brightest emission to a flatter component ($\alpha$=$-$0.30$\pm$0.07) in the surrounding fainter region.
To probe the interstellar medium into which the redefined SNR is likely evolving, we have analysed the surrounding atomic and molecular gas using the 21~cm neutral hydrogen (HI) and $^{13}$CO (J=1$-$0) emissions. 
We found that G51.04+0.07 is confined within an elongated HI cavity and that its radio emission is consistent with the remains of a stellar explosion that occurred $\sim$6300~yr ago at a distance of 7.7$\pm$2.3~kpc. Kinematic data suggest that the newly discovered SNR lies in front of HII regions in the complex, consistent with the lack of a turnover in the low frequency continuum spectrum. The CO observations revealed molecular material that traces the central and northern parts of G51.04+0.07. 
The interaction between the cloud and the radio source is not conclusive and motivates further study. The relatively low flux density ($\sim$1.5 Jy at 1400 MHz) of G51.04+0.07 is consistent with this and many similar SNRs lying hidden along complex lines of sight towards inner Galactic emission complexes. It would also not be surprising if the larger complex studied here hosted additional SNRs.}


\maketitle

\section{Introduction}
\label{intro}
Supernovae (SNe) explosions and their remnants (SNRs) play a major role in key areas across Galactic astrophysics. As the presumed source of Galactic cosmic rays, they seed the interstellar medium (ISM) with at least approximately one third of its energy density and drive its structure and evolution \citep{mck77}, including the stimulation of new generations of star formation. The rate of massive star formation and the Galactic SNe rate must also be related. These and other factors motivate an accurate census of supernova remnants as key tracers of both stellar evolution and SNe. 
Despite considerable effort, the number of identified Galactic SNRs, $\sim$300 \citep{gre17}, remains significantly lower than expected due to severe observational selection effects. Indeed, statistical studies, including OB stars counts, pulsar birth rates and supernova rates in other Local Group galaxies predict that there should be approximately one to two stellar explosions per century in our Galaxy \citep{li91, tam94}. 
For example, by assuming that the radio synchrotron emission from an SNR is observable for 50,000 to 100,000 years, the number of Milky Way SNRs should be at least three times greater than currently known. The same conclusion on the Galactic remnant's population had already been reached by \citet{helfand+89} on the basis of the number and distribution of SNRs in the Galaxy by then catalogued  \citep{green-cat-88}.

The evidence is clear that the `missing' SNRs include both the youngest and oldest SNRs. The former were missed by the early generations  of single-dish Galactic plane surveys that identified SNRs based on their extended morphology but at limited angular resolution. The latter, the older extended SNRs, were also missed due to limited dynamic range in surface brightness sensitivity, leaving them confused along complex Galactic sight-lines including the superposition of thermal and nonthermal sources of varying sizes and brightnesses. 
For several decades now, observations have proceeded painstakingly, uncovering a few new SNRs per year, at best \citep[see for instance,][]{bro04,tia08, roy13, sab13, gre14, kot17}. A significant step forwards filling out the census of Galactic SNRs was achieved with the 330~MHz system on the legacy Very Large Array (VLA) \citep{bro06snr35}. Together with a rich archive of higher frequency radio and infrared observations to distinguish them from HII regions, this 330~MHz interferometric survey uncovered 35 new SNRs in a relatively limited region of the inner Galaxy. 
This work illustrated the advantage of lower frequency interferometer observations in overcoming selection effects that have hidden many SNRs, and suggests a resurgence in higher angular resolution, low-frequency observations could impact the field significantly. Newer surveys like GLEAM \citep{way15,hur17} could be particularly potent as they probe into regions of the inner Galaxy unreachable from northern facilities.

In this paper, we present compelling evidence for a new Galactic supernova remnant, named \object{G51.04+0.07}. The unambiguous identification of its nonthermal radio continuum emission was revealed at 74~MHz as part of the Very Large Array Low-frequency Sky Survey Redux \citep[VLSSr,][]{lan14}. These data are used in conjunction with available higher-frequency radio images to constrain the continuum spectrum of the source. We have also investigated the properties of the ISM into which the SNR~G51.04+0.07 is evolving, by using public datasets from the 21~cm VLA Galactic Plane Survey \citep[VGPS,][]{sti06} and $^{13}$CO (J=1$-$0) observations from the Boston University-Five College Radio Astronomy Observatory Galactic Ring Survey \citep[GRS,][]{jac06}.

\section{Results}
\subsection{The radio continuum emission: morphology and spectrum}
\label{radio}
Recently, \citet{and17} presented a list of new  SNR candidates in the Galaxy identified on the basis of 1-2~GHz continuum data from The HI, OH, Recombination line survey of the Milky Way (THOR) combined with those from the VLA 1.4~GHz Galactic Plane Survey (VGPS) and infrared survey observations. The large ($\sim$30$^{\prime}$), complex Galactic region, \object{G51.21+0.11}, with an integrated flux density at 1400~MHz of 24.35~Jy, is part of their compilation of candidates. 
Our observations, including new data at 74~MHz, have allowed us to take the next step in disentangling the superposition of thermal and nonthermal emission in the complex. As a result, we ascertain that the complex likely contains, not unexpectedly, multiple SNRs and HII regions. In this paper we focus on a region of nonthermal emission that we define as the SNR G51.04+0.07, which is considerably smaller than the original classification of the larger SNR candidate made by \citet{and17}. We note that there is evidence of nonthermal emission elsewhere in the complex, indicating the possible presence of at least one more SNR. 

\begin{figure}[!ht]
\centering
\includegraphics[width=8cm]{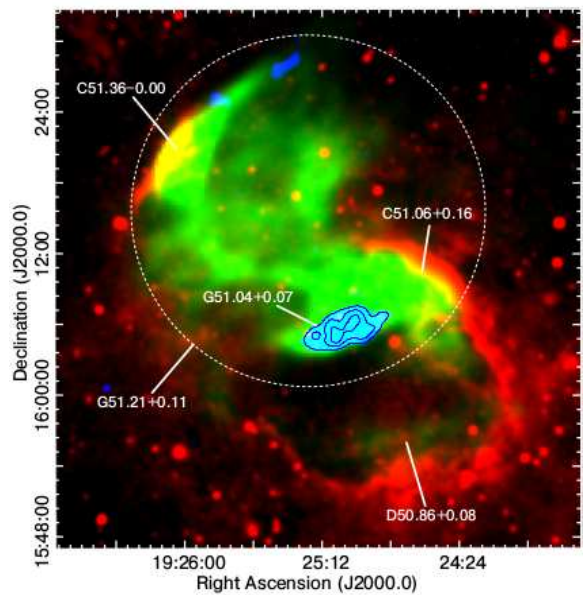} 
\caption{Colour composite image showing the spatial correspondence between the mid-infrared emission as observed by {\itshape MSX} at 8~$\mu$m (in red), and the radio continuum emission at 74 (in blue) and 1420~MHz (in green) from the VLSSr and VGPS, respectively. Thermal emission features are traced in yellow and green, while nonthermal radiation is visible in cyan. The overlaid contours tracing the 74~MHz nonthermal emission from the new SNR G51.04+0.07 are at 0.34, 0.58, and 0.78~Jy~beam$^{-1}$. The dotted circle indicates the size and position of the previously proposed SNR candidate G51.21+0.11 \citep{and17}. The HII regions in the field are also labelled.}
\label{ir-radio}
\end{figure}

Figure~\ref{ir-radio} shows a three-colour map of the region corresponding to the source G51.21+0.11. The image compares the infrared {\it Midcourse Space Experiment} image at 8~$\mu$m (red) \citep[{\itshape MSX},][]{pri01} with superimposed total intensity radio-continuum  images from  the VLSSr and VGPS at 74~MHz (blue) and 1420~MHz (green), respectively. 
The image at the highest radio frequency combines data from the D configuration of the VLA with Effelsberg single dish observations \citep{sti06}.%
\footnote{Because the area located to the north-east of G51.04+0.07 is not covered by the `THOR+VGPS' data \citep{beu16}, we only used the observations provided by the VGPS to produce the multi-frequency comparison presented in Fig.~\ref{ir-radio}.} 
The {\itshape MSX}-8~$\mu$m image traces thermal gas emission from polycyclic aromatic hydrocarbons, characteristic of HII regions. 
At 1420~MHz both thermal and nonthermal emission components are detected. Nevertheless, at such low frequencies as that of the VLSSr nonthermal features dominate the band. 
Within this context, the information provided by the 74~MHz data from the VLSSr is unique for spatially segregating the thermal and nonthermal emission and constraining the morphology and flux density of the SNR candidate emission. In the field of view shown in Fig.~\ref{ir-radio}, a distinctive cyan feature approximately at the center of the field, where both the emission at 74 and 1420~MHz positionally coincide, denotes a region where purely nonthermal emission occurs. 
Although the VLSSr has limited sensitivity to diffuse structures on 30$^{\prime}$ scales, we would still expect to detect smaller emission knots and the s-shaped ridge of emission visible in the VGPS image if that emission were nonthermal. 
The yellow structures in our colour representation trace thermal regions where the radio emission at 1420~MHz and the mid-infrared radiation overlap. The agreement between these two spectral bands observed to the north-east and west in the field of view corresponds to the photodissociation regions associated with the HII regions \object{C51.36$-$0.00} (centered at R.A.$\sim$$19^{\mathrm{h}}\,26^{\mathrm{m}}\,00^{\mathrm{s}}$, Dec$\sim$$16^{\circ}\,20^{\prime}\,00^{\prime\prime}$, J2000.0) and \object{C51.06+0.16} (centered at $\sim$$19^{\mathrm{h}}\,24^{\mathrm{m}}\,48^{\mathrm{s}}$, $\sim$$16^{\circ}\,09^{\prime}\,00^{\prime\prime}$) in the catalogue presented by \citet{and09}. 
Regarding the HII region \object{D50.86+0.08} ($\sim$$19^{\mathrm{h}}\,24^{\mathrm{m}}\,42^{\mathrm{s}}$, $\sim$$15^{\circ}\,56^{\prime}\,00^{\prime\prime}$), we note that 
because of the low surface brightness observed at 1420~MHz it does not appear as a bright yellow feature (see Fig.~\ref{ir-radio}). 
For this and the C51.06+0.16 HII region, \citet{and09} derived kinematical distances of $\sim$8~kpc. This was based on recombination-line emission detected at 42.9$\pm$1.2~km~s$^{-1}$ and 42.2$\pm$0.5~km~s$^{-1}$, respectively, with the NRAO Green Bank 140-foot telescope, together with existing sky surveys of neutral hydrogen (HI) and $^{13}$CO.\footnote{As described by \citet{and09}, uncertainties up to 2~kpc in their analysis can result from velocity deviations of $\sim$10~km~s$^{-1}$ caused by the streaming motions in the HI gas.} 
Although no radio recombination line detection towards the HII region C51.36$-$0.00 has been reported in the literature, the catalogued distance to this thermal source is $\sim$6.4~kpc \citep{and09}. 
We also note that in the surveyed area, two sites of massive star formation \citep{lum13} as well as spots of H$_ {2}$O and CH$_3$OH maser emission \citep{cod94,szy12} were identified. Finally, as alluded earlier, marginal evidence for nonthermal emission at the north-east of Fig.~\ref{ir-radio} may indicate the presence of at least one more SNR in the larger complex. 

\begin{figure}[!ht]
\centering
\includegraphics[width=8cm]{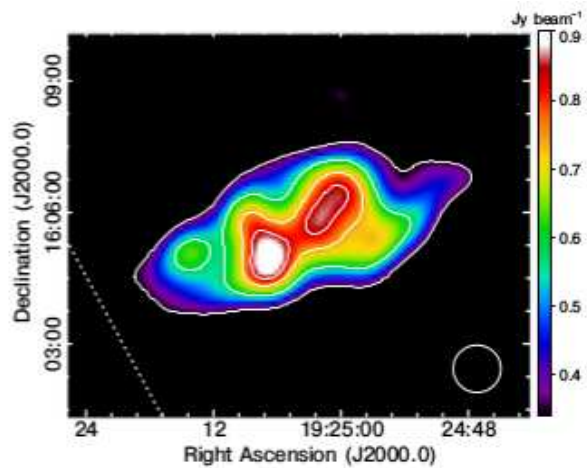}
\caption{A close view of the newly identified SNR~G51.04 at 74~MHz from the VLSSr. This image is strong evidence for the nonthermal nature of the emission and is currently the only available image of this source below 100~MHz. The measured sensitivity level is 0.095~Jy~beam$^{-1}$, and the contours are traced at 0.34, 0.48, 0.78, and 0.83~{\Jyb}. The synthesized beam shown at the bottom right corner is 75$^{\prime\prime}$$\times$75$^{\prime\prime}$. The position and direction of the Galactic plane are indicated by the dotted line.}
\label{radio74}
\end{figure}

Even though it was not previously recognised as a distinct zone, an elongated nonthermal structure is now evident in the field. In Fig.~\ref{radio74} we present a zoomed view of this region, underscored by its emission at 74~MHz. As mentioned above, we refer to this region as the source G51.04+0.07 (hereafter, G51.04), according to its approximate geometric center. In what follows, we provide morphological and spectral evidence of the nature of this source as a new SNR in our Galaxy. Clearly the 74~MHz emission identifying G51.04 is significantly less  extended than G51.21+0.11, the latter being a composite of both thermal and nonthermal emission, although previously classified by \citet{and17} as a single SNR. This is not necessarily surprising, as a mixture of thermal and nonthermal radio emission in such area was mapped by the THOR+VGPS data.

From the low-frequency radio image, the emission from G51.04 covers an elliptical region with major and minor axes of \dms{7}{\prime}{5}$\times$3$^{\prime}$ in the south-east and north-west direction. The large aspect ratio of the source ($\sim$2.5:1) might be associated with an inhibited expansion in the direction of the Galactic plane. However, no significant brightening is observed in this direction, as one might expect due to enhanced compression. 
With an angular resolution of 75$^{\prime\prime}$ and a sensitivity of 0.095~Jy~beam$^{-1}$, the image at 74~MHz reveals that the brightest emission in G51.04 occurs, at least as we see projected on the plane of the sky, inside the radio shell.
A maximum with brightness 0.15~Jy~beam$^{-1}$ located at $\sim$19$^{\mathrm{h}}$25$^{\mathrm{m}}$07$^{\mathrm{s}}$, $\sim$16$^{\circ}$05$^{\prime}$ is noticeable.

We considered, but disfavour, an interpretation of the bright central feature as a pulsar driven nebula (PWN) for one reason: its continuum spectrum ($\alpha$$\sim$$-$0.5, see below) is too steep, and PWNs normally have relatively flat spectra. Moreover, no radio pulsar coincident with G51.04 has been detected in the latest ATNF Pulsar Catalogue%
\footnote{\url{http://www.atnf.csiro.au/research/pulsar/psrcat/}} \citep{man05}, version 1.57 (accessed 15 November, 2017), with the nearest pulsar, PSR~J1926+1613, located 15$^{\prime}$ away. Our search also yields negative results in other pulsar surveys.%
\footnote{See for instance \url{http://www.ioffe.ru/astro1/psr-catalogue/Catalog.php} and \url{https://confluence.slac.stanford.edu/display/GLAMCOG/Public+List+of+LAT-Detected+Gamma-Ray+Pulsars}.} 
A fainter radio spot (${\sim}$0.1~Jy~beam$^{-1}$) to the north-west near $\sim$19$^{\mathrm{h}}$25$^{\mathrm{m}}$02$^{\mathrm{s}}$, $\sim$16$^{\circ}$06$^{\prime}$ is also resolved in the low-frequency image. The brightness gradually decreases in strength towards the edges of the source. The boundaries of G51.04 are not particularly sharp. 

The integrated flux density of G51.04 at 74~MHz is 6.1$\pm$0.8~Jy. We computed the error in this estimate by taking several sources of uncertainty into account (added in quadrature), the main of which are the ones associated with intrinsic errors in the calibration of the absolute flux-density scale ($\sim$15\%), clean bias errors modified for extended sources, and uncertainties in defining the outer boundary of the radio emission.

\begin{figure}[!ht]
 \centering
\includegraphics[scale=0.7]{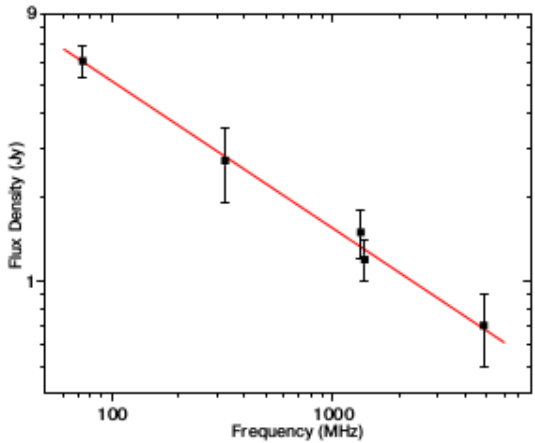}
\caption{Global radio continuum spectrum of G51.04 obtained from the flux density measurements summarised in Table~\ref{fluxes}. The straight line represents a weighted fit performed to all the plotted values with a power-law index $\alpha$=$-$0.52$\pm$0.05.}
\label{spectrum}
\end{figure}

On the basis of the flux density measurement at 74~MHz for G51.04 combined with higher frequency measurements from radio surveys reported in Table~\ref{fluxes}, we estimated the radio continuum spectrum of the source. All the measured flux densities were tied to the flux scale of \citet{per17}. In Fig.~\ref{spectrum} we show a plot of the integrated spectrum for G51.04. A weighted fit to the data points using a single power law yields a slope $\alpha$=$-$0.52$\pm$0.05 (defined by the relation $S_{\nu} \propto \nu^{\alpha}$, with $S_{\nu}$ the flux at the frequency $\nu$). This result shows conclusively that the radiation from G51.04 has a nonthermal origin. Moreover, the derived spectrum is entirely in concordance with spectral indices measured in many other recognised shell-type Galactic SNRs \citep[for which the mean spectral index is $\alpha$$\sim$$-$0.5,][]{kov94}. It is noteworthy that in spite of the thermal emission present in the larger G51.21+0.11 Galactic complex, the spectrum of G51.04 does not show evidence of a turnover at frequencies $\nu$<100~MHz. 
Low frequency spectral turnovers towards Galactic SNRs are common but not ubiquitous, and are attributed to an inhomogeneous component of low density ionised gas surrounding normal HII regions along the line of sight \citep{kas89}. Absorption from ionised gas has now been spatially resolved by the VLA at 74~MHz, towards, for example, the SNRs W49B \citep{lac01}, 3C~391 \citep{bro05}, and IC~443 \citep{cas11}. In the two latter cases the absorption indicates the SNRs are clearly interacting with a surrounding molecular cloud complex. The lack of clear absorption towards G51.04 at 74~MHz might indicate that it is in the foreground relative to the larger G51.21+0.11 complex. 

Using the integrated flux density estimate and the angular extent measured at 74~MHz, the surface brightness of the newly discovered SNR~G51.04, $\Sigma$=1.5$\times$10$^{-19}$ ($S$/$\theta$$^{2}$)~W~m$^{-2}$~Hz$^{-1}$~sr$^{-1}$, where $S$ and $\theta$, respectively, represent the flux (in Jy) and the angular size (in arcmin) of the source, is $\Sigma_{\mathrm{74 MHz}}$$\simeq$ 4.1$\times$ 10$^{-20}$~W~m$^{-2}$~Hz$^{-1}$~sr$^{-1}$. 
For a reference frequency of 1~GHz, we obtained $\Sigma_{\mathrm{1 GHz}}$ $\simeq$ 1 $\times$ 10$^{-20}$~W~m$^{-2}$~Hz$^{-1}$~sr$^{-1}$. This result is consistent with the range presented by \citet{pav13}, encompassing values 10$^{-21}$-10$^{-19}$~W~m$^{-2}$~Hz$^{-1}$~sr$^{-1}$ at 1~GHz for a large sample Galactic SNRs. This surface brightness ranks G51.04 at about thirty out of $\sim$180 objects classified as shell-like SNRs \citep{gre17}. This suggests that G51.04 is one of the high surface brightness SNRs known to missing from catalogues due to observational selection effects. 
By integrating the spectrum from 10$^{7}$ to 10$^{11}$~Hz, we derived a radio luminosity of $\sim$2$\times$10$^{33}$~erg~s$^{-1}$ for a distance to the source of $\sim$7.7~kpc (see Sec.~\ref{HICO}), consistent with typical radio luminosities of Galactic SNRs \citep[e.g.][]{ilo72,pav13}. Taken together, the derived radio properties for G51.04 are all consistent with a supernova remnant origin.

\begin{table*}[h!]
\centering
\caption{Flux densities used to construct the global radio continuum spectrum of G51.04.}
\label{fluxes}
\begin{tabular}{ccl}\hline\hline
 Frequency &   \multicolumn{1}{c}{Flux density}  & \multicolumn{1}{c}{\multirow{2}{*}{Survey}}                     \\
   (MHz)   &       \multicolumn{1}{c}{(Jy)}      &                                                \\\hline
 ~~~~74    &             6.1 $\pm$ 0.8           &  VLSSr \citep{lan14}                           \\
  ~~327    &             2.7 $\pm$ 0.8           &  Westerbork Northern Sky Survey \citep{ren97}  \\
   1400    &             1.5 $\pm$ 0.3           &  THOR+VGPS \citep{beu16}                       \\ 
   1400    &             1.2 $\pm$ 0.2           &  NRAO VLA Sky Survey  \citep{con98}            \\
   4850    &             0.7 $\pm$ 0.2           &  4.85 GHz sky survey \citep{con94}             \hfil\\\hline
\end{tabular}
\end{table*}

\begin{figure}[!ht]
\centering
\includegraphics[width=8cm]{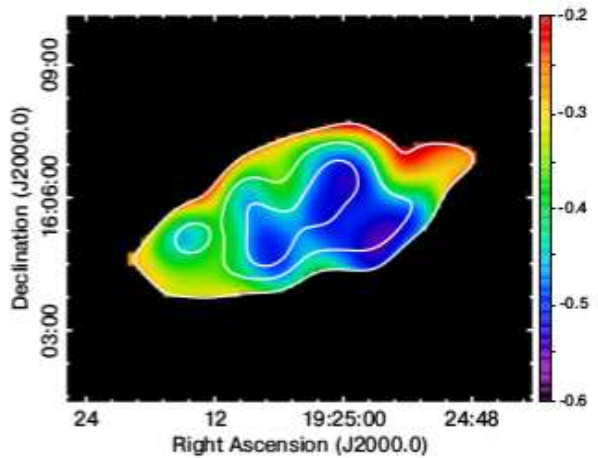}
\caption{Spatial spectral index distribution in the SNR~G51.04 calculated between 74 and 1400~MHz using the VLSSr and THOR+VGPS images. To create this spectral map, which has a resolution of 75$^{\prime\prime}$, the input images at 74 and 1400~MHz were masked at 3.5$\sigma$ and 3$\sigma$ of their respective noise levels. The superimposed contours correspond to the 74~MHz emission at levels 0.34, 0.58, and 0.78~Jy~beam$^{-1}$.}
\label{alphamap}
\end{figure}

In addition to the global radio continuum analysis, we made spatially resolved spectral measurements searching for trends in the spectral index across G51.04. To accomplish this we used the VLSSr image at 74~MHz together with the image at 1400~MHz from THOR+VGPS. We notice that, because the visibility data were not available for the image at the higher frequency, this dataset was not limited in its \it uv \rm range nor tapered to match the dirty beam at 74~MHz. Hence, the spectral index map shown in Fig.~\ref{alphamap} was constructed by convolving the highest frequency image to the 75$^{\prime\prime}$ synthesized beam of the VLSSr-74~MHz image. 
Before calculating spectral indices, the images at 74 and 1400~MHz were also aligned, interpolated to identical projections, and clipped at a 3.5$\sigma$ and 3$\sigma$ significance level of their respective sensitivities (at 1400~MHz the rms noise level and the synthesized beam are $\sim$5~mJy~beam$^{-1}$ and 25$^{\prime\prime}$, respectively).
The errors in our determination of the spectral index from the map are 0.03 - 0.09 for the brightest and the faintest regions, respectively.
In addition to the results presented here, we emphasise that a good quality image at an intermediate frequency between 74 and 1400~MHz is needed to confirm and improve upon the spectral characteristics of G51.04 derived here. 

Figure~\ref{alphamap} displays the spatial distribution of the spectral index in the redefined SNR obtained from the direct ratio of the 74 and 1400~MHz images. The spread in the spectral indices is of the order of $\pm$0.34 across the remnant. A steep spectral index region, with an average value $\alpha_{74}^{1400}$$\sim$$-$0.5, is easily identified in this map, which spatially corresponds to the brightest central region in G51.04. The map also reveals a clear trend of a flattening up to $\alpha_{74}^{1400}$$\sim$$-$0.3 towards the surrounding lower brightness region. This is most evident in the northwestern border of G51.04. 
One possibility is that this region is flattened due to the superposition of thermal emitting gas contaminating the remnant's spectral indices, which could be important in the higher frequency image used to construct the spectral map.
Overall, we note that the spectral components do not appear to  be correlated with features in total intensity, in the sense that the brighter regions are not those with flatter spectrum as is expected under the first-order particle acceleration mechanism. A similar spectral behaviour has been found in the shell-class Galactic SNRs G39.2$-$0.3, G41.1$-$0.3 \citep{and93}, and G15.4+0.1 \citep{sup15}, for which it was proposed that different mechanisms are responsible for the spectral and brightness variations observed across these sources.

\subsection{Distance and age determinations}
\label{dist-age}
\begin{figure}[!ht]
\centering
\includegraphics[width=8cm]{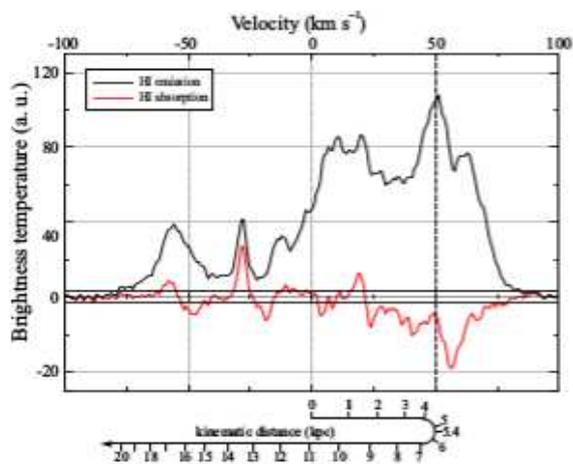} 
\caption{ 21~cm emission (black curve) and absorption (red curve) spectra towards the continuum source G51.04. The dashed vertical line indicates the velocity of the tangent point. The horizontal solid lines correspond to the rms noise level of 3.2~K in the absorption profile of the HI. Below, the assigned kinematic distances are denoted \citep{fic89}.}
\label{HIdistance}
\end{figure}

In the following, we attempt to put constraints on basic parameters for G51.04 such as its age and distance. For the distance estimate we used 21~cm neutral hydrogen absorption measurements from the VGPS. In Fig.~\ref{HIdistance} we show the HI emission spectrum constructed from an area projected against the brightest central portion of the SNR~G51.04. The absorption profile is also included in the figure. It was obtained by subtracting the HI emission spectrum from a profile averaged over different patches devoid of continuum emission. 
Significant and continuous absorption is observed in the velocity range between 20~km~s$^{-1}$ and that of the tangential point ($v_{\mathrm{TP}}$$\sim$50~km~s$^{-1}$) (all the radial velocities mentioned in this paper are measured with respect to the local standard of rest, LSR). 
The most positive velocity HI absorption feature at $\sim$56~km~s$^{-1}$ likely corresponds to gas at the velocity of the tangential point and could be caused by the expanding motions of the HI gas due to the SNR shockwave. 
According to the circular rotation model of \citet{fic89} (with the Galactocentric distance R$_ {0}$=8.5~kpc and the rotation velocity at the Sun $\Theta$=220~km~s$^{-1}$), 
the velocity of the tangential point corresponds to a lower limit on the distance of 5.4~kpc. At negative velocities a weak absorption feature near $-$19~km~s$^{-1}$ is observed. However, in view of its association with relative low amplitude emission we regard this feature as unreliable.
The same argument holds for the low-level absorption peak at $-$50~km~s$^{-1}$. In this case, the non-detection of absorption in correspondence with the emission feature in the HI spectrum centered at $-$30~km~s$^{-1}$ is interpreted as additional support of our suggestion that the feature at the most negative velocity is not genuine. Certainly, if it were real there should be absorption coincident with the emission at $-$30~km~s$^{-1}$, since there is no kinematic distance ambiguity (or velocity blending) for the gas in this direction of the Galaxy moving at negative velocities. Therefore, in light of the results presented here, an upper limit of $\sim$11~kpc on the kinematic distance to G51.04 can be established, which approximately corresponds to the radial velocity of 0~km~s$^{-1}$. 
We present in Sec.~\ref{HICO} a further detailed discussion of this estimate, on the basis of the HI distribution in the direction to G51.04. 

For the purpose of deriving the age of G51.04 we note that its angular size suggests it is not an evolved object, a condition that is not particularly constraining. So far, the region has not been observed in X rays. In examining optical datasets (see for instance, SuperCOSMOS H-alpha Survey, \citealt{par05} and The STScI Digitized Sky Survey%
\footnote{\url{http://archive.stsci.edu/dss/}.}),
we did not find obvious correlations with G51.04. 
To derive an age, we have assumed that the source is in the Sedov-Taylor phase of its evolution. In this evolutionary stage, the age of the remnant is given roughly by $0.4R_{\mathrm{ST}}/v$, the ratio between the radius and the expansion velocity of the blast wave, being $R_{\mathrm{ST}} \simeq 14.5\,(E_{\mathrm{SN}}/10^{51}\,\mathrm{erg})^{1/5}\,(0.5\,\mathrm{cm}^{-3}/\eta_{0})^{1/5}\,t_{4}^{2/5}$~pc, for a shock with a specific heat ratio $\gamma$=5/3 \citep{tie05}. Here, $E_{\mathrm{SN}}$ is the energy of the supernova, $\eta_{0}$ is the number density of hydrogen, while $t_{4}$ is the age of the remnant normalised to 10$^{4}$~yr. 
Therefore, using a measured mean density of atomic hydrogen in the ambient $\eta_{0}$$\sim$10~cm$^{-3}$ (as we explain later in Sec.~\ref{HICO}), we calculated an age for the SNR of about 6300~yr. It represents an average between the ages corresponding to the major and minor semiaxis of G51.04 (\dms{3}{\prime}{75}$\times$\dms{1}{\prime}{5}, or 8.4 and 3.4~pc calculated at a distance of 7.7$\pm$2.3~kpc, see Sec.~\ref{HICO}). 
We emphasise that this is a very rough estimate due to the uncertainties in the measurements involved and the assumption of the Sedov stage in the SNR's evolution.

\subsection{HI and CO environments}
\label{HICO}
\begin{figure}[!ht]
\centering
\includegraphics[width=8cm]{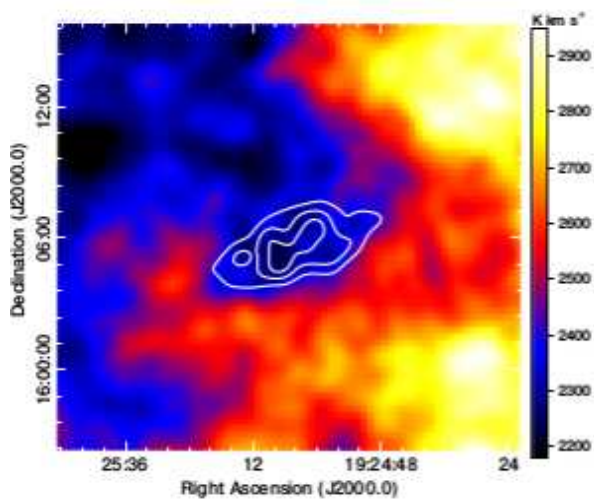}
\caption{21~cm line emission in the direction of G51.04. The map corresponds to the integration of the neutral gas from 25 to 55~km~s$^{-1}$. The 0.34, 0.58, and 0.78~Jy~beam$^{-1}$ contours from the 75$^{\prime\prime}$ resolution 74~MHz VLSSr image are included to facilitate the comparison between G51.04 and its surroundings.}
\label{HIcavity}
\end{figure}

To analyse the environs of G51.04, we used observations of the neutral gas emission in the 21~cm line and $^{13}$CO (J=1$-$0) molecular data from the VGPS and the GRS, respectively. 
The inspection of the atomic gas throughout the range of velocities covered by the VGPS strikingly shows, in the radial velocity range from $\sim$25 to $\sim$55~km~s$^{-1}$, a distortion in the HI distribution at the position of the synchrotron emission from G51.04. The distortion resembles a cavity-like structure in the HI emission, which looks open towards the northern border of G51.04. This feature is presented in Fig.~\ref{HIcavity}, where contours of the radio emission from the VLSSr-74~MHz image are included for reference. 
To construct this map, an appropriate background level equal to the mean value of each HI image was subtracted. Based on its morphology, we suggest that the HI cavity has possibly been caused by the expansion of the blastwave associated with G51.04 into the HI surroundings. We are aware that similar HI structures have been recognised towards a number of Galactic SNRs, and were interpreted as wind-blown bubbles created by the massive SN progenitor \citep[see for instance,][]{max13}. 
However, results obtained from models and observations have also demonstrated that a scenario in which a cavity-like distribution carved by an SNIa progenitor is also plausible \citep{bad07,wil11}. Nevertheless, we highlight here that, based on radio data, we did not find evidence of a PWN inside G51.04.

If we assign a central velocity of $\sim$40~km~s$^{-1}$ for the HI cavity, the corresponding near and far kinematic distances are 3.3 and 7.7~kpc \citep{fic89}. Based on the hypothesis of association between the cavity-like feature and the source G51.04 (for which we determined from HI absorption measurements a range of 5.4-11~kpc), we consider the value of 7.7$\pm$2.3~kpc as the most plausible distance for the detected HI structure and the nonthermal radio source G51.04 related to it. The error in this determination includes uncertainties inherent to the circular rotation model and the ones related to the determination of the radial velocity of the HI signature identified in association with G51.04. 
To complete this picture, we also recall that both the distance and systematic velocities measured for the HII regions in the field of view around G51.04 ($v_{\mathrm{RRL}}$$\sim$42~km~s$^{-1}$, $d$$\sim$8~kpc, \citet{loc96}; also see our Fig.~\ref{ir-radio}) are similar to the distance determined here for both the HI cavity and the SNR G51.04. 
Therefore, we conclude that all the observed features, thermal and nonthermal, and the HI are related. Furthermore, taking the lack of a low frequency continuum spectral turnover into account (as discussed in Sec.~\ref{radio}), the SNR~G51.04 is closer than the HII regions in the complex, or just on their near side.

In determining the physical parameters of the neutral gas that we assume is associated with G51.04, we adopted for the HI cavity a center $\sim$$19^{\mathrm{h}}\,25^{\mathrm{m}}\,05^{\mathrm{s}}$, $\sim$$16^{\circ}\,05^{\prime}\,44^{\prime\prime}$, a radius of 6$^{\prime}$$\times$\dms{3}{\prime}{5} and a thickness of \dms{2}{\prime}{5} (equal to $\sim$13.4$\times$7.8~pc and 5.6~pc at a distance of 7.7~kpc). 
Under the assumption that the HI emission is optically thin, the HI column density is given as $N_{\mathrm{H}}=1.823 \times 10^{18} \, \int{T_{\mathrm{b}} \, \mathrm{d}v}$, where $T_{\mathrm{b}}$[K] and $v$[km~s$^{-1}$] denote the intensity and the velocity of the neutral gas, respectively \citep{dic90}. 
We thus calculated the average column density for the cavity-like structure to be $N_{\mathrm{H}}$=0.74$\times$10$^{21}$~cm$^{-2}$. We note that this value could be interpreted as an upper limit due to possible confusion with unrelated foreground and background gas.
We also found that, at a distance of 7.7~kpc, the obtained column density implies an average density in the HI cavity of about $\eta$$\sim$$N_{\mathrm{H}}/L$=23~cm$^{-3}$, using a characteristic length of the cavity equal to the mean radius of the HI feature, \dms{4}{\prime}{75} (or 10.6~pc at $d$$\sim$7.7~kpc). 
Therefore, if the neutral gas was originally distributed inside a cylinder of radius and line-of-sight depth of 10.6~pc, the derived mean density of hydrogen atoms in the medium is $\eta_{0}$$\sim$10~cm$^{-3}$. 
This estimate is fairly consistent with the 20-50~cm$^{-3}$ hydrogen density in cold HI clouds \citep[e.g.][]{dra98,fer01}. The mean relative error in the determination of parameters for the neutral gas is about 40\%, stemming mainly from uncertainties in the distance determination, in the selection of integration boundaries, as well as the inclusion of possible unrelated gas located along the line of sight.

\begin{figure*}[!ht]
\centering
\includegraphics[width=0.75\textwidth]{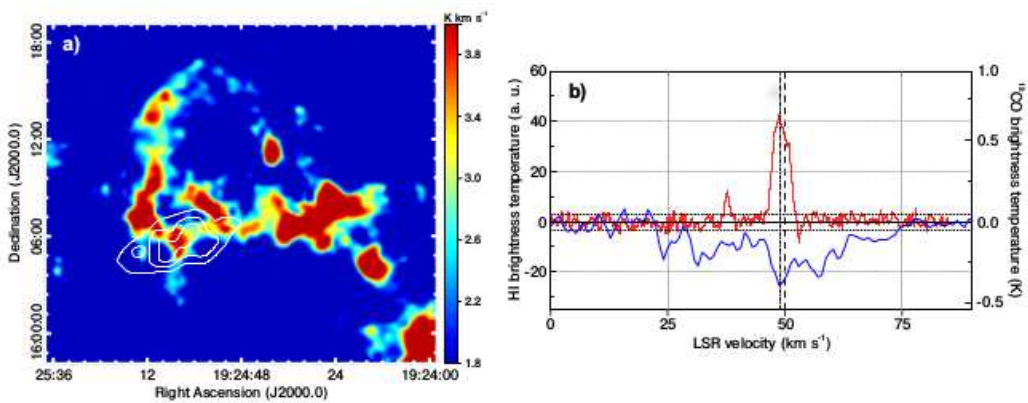}
\caption{ {\bfseries a)} Emission of $^{13}$CO (J=1$-$0) in the Galactic complex associated with the source G51.04 and the star forming activity regions shown in Fig.~\ref{ir-radio}. The image was obtained by integrating the molecular line emission in the velocity range from 40 to 52~km~s$^{-1}$. The contours trace the 74~MHz radio continuum emission from G51.04 at 0.34, 0.58, and 0.78~Jy~beam$^{-1}$. 
{\bfseries b)} $^{13}$CO emission (red curve) and HI absorption (blue curve) spectra  towards the identified molecular component, which are roughly coincident in projection with the radio emission from G51.04. 
Horizontal dotted lines mark the rms noise of the HI absorption spectrum. The dash-triple-dot line indicates the velocity of the cloud, while the dashed line marks the 
velocity of the tangent point. The spectral features allow us to establish a lower kinematic distance  limit of $\sim$6~kpc to the cloud.}
\label{CO}
\end{figure*}

We have also searched for molecular structures traced by the $^{13}$CO (J=1$-$0) gas related to G51.04. After inspecting  the entire data cube from the GRS in the velocity range between $-$5 and 135~km~s$^{-1}$, we identified a molecular cloud which seems to match (in projection) the radio emission from G51.04. The integrated emission from this cloud in the LSR range $\sim$40-52~km~s$^{-1}$ is depicted in Fig.~\ref{CO}a, where contours of the 74~MHz radio emission from VLSSr image are included for ease of comparison. 
The question to be addressed now is whether this molecular material is part of the complex environment associated with G51.04 and the HII regions. To analyse this point we used the 21~cm continuum absorption method presented by \citet{rom09} as a tool to determine the distance to the molecular gas. 
The $^{13}$CO average spectrum for the cloud is shown in Fig.~\ref{CO}b. It was extracted from a $\sim$\dms{1}{\prime}{5} box centered at the brightest part of the cloud, that, seen in projection, overlaps the continuum emission from G51.04. Figure~\ref{CO}b also includes the HI absorption profile towards the cloud. It was obtained by subtracting the HI emission profile from a spectrum averaged over several regions free from $^{13}$CO emission. At least two peaks are evident in the $^{13}$CO emission line spectrum, one at $\sim$37~km~s$^{-1}$ and the other at $\sim$49~km~s$^{-1}$. 
The latter, very close to the velocity of the tangent point, corresponds to the molecular material detected in the direction of G51.04. A correlation between these CO peaks and HI absorption features is also noticeable. Because the HI gas is ubiquitously distributed in the Galaxy, at the smaller radial velocity this correlation would be produced by foreground molecular material absorbing the continuum radiation from G51.04. 
Regarding the 49~km~s$^{-1}$ emission line peak, we are aware that if the cloud containing the radio continuum source is at the far kinematic distance ($\sim$6~kpc) then the correspondence of HI absorption features with CO emission peaks should be observed at velocities greater than the velocity of the cloud of interest (up to the velocity of the tangent point). However, this picture becomes confusing in the case of the discovered cloud. 
Firstly, although the 49~km~s$^{-1}$ emission line peak appears to be broadened up to the velocity of the tangent point, as shown in Fig.~\ref{CO}b, it is difficult to determine whether this feature represents real perturbations of the molecular gas or simply a multiplicity of velocity components along the line of sight. 
Secondly, CO emission features and the corresponding absorption in the 21~cm continuum are observed in Fig.~\ref{CO}b at a velocity greater than the velocity of the tangent point, but this probably originates in random motions of the gas, that locally can produce velocities greater than those corresponding to the tangent point \citep{rom09}. 
Therefore, taking all the spectral signatures together, we established a lower limit to the kinematic distance to the cloud at 49~km~s$^{-1}$ of $\sim$6~kpc, the far distance assigned to this radial velocity \citep{fic89}.

We now use our analysis of the molecular gas distribution in conjunction with both the morphological and spectral features observed across G51.04. According to our interpretation of the spectra presented in Fig.~\ref{CO}b, a physical association between the cloud and the radio source G51.04 should not be dismissed (within the uncertainties in the determination of the distances to G51.04 and the cloud). 
Consequently, as a possibility, the  brightest part in the 74~MHz radio emission of G51.04, that we see projected on the plane of the sky lying at the center of the shell, may be in fact indicating a region where the SN shock front is just encountering part of the molecular gas. This does not necessarily mean a critical interaction of G51.04 with its surroundings. 
The local synchrotron spectrum of G51.04  in this region ($\alpha$$\sim$$-$0.5) is compatible with a first-order Fermi mechanism operating at the shock. It is also interesting to notice that there is CO molecular gas in apparent coincidence with the localised spectral index flattening observed in the northwestern edge of G51.04. Such a correspondence could be due to a shock that is being decelerated while it is running through the high density interstellar material. Although this component has a spectrum appreciably flatter than the surrounding synchrotron plasma, no spectral index signature suggesting the presence of thermal absorption from collisional ionisation was found in the postshock region \citep{bro05,cas11}. 
Of course, we recognise that an alternative scenario in which thermal gas associated with the neighbouring HII regions determines the flattening cannot be ruled out. More sensitive radio continuum data is needed before advancing a detailed interpretation of the local spectral variations and its relationship with the observed CO features.

Finally, we point out that an excess of $\gamma$-ray emission at GeV energies (catalogued as FL8Y~J1925.4+1616%
\footnote{The preliminary {\Fermi} 8-year Point Source List (FL8Y), which will be soon superseeded by the forthcoming {\Fermi} Fourth Source Catalog (4FGL), is available at \url{https://fermi.gsfc.nasa.gov/ssc/data/access/lat/fl8y/}.})
has been detected with the Large Area Telescope onboard the {\itshape Fermi} satellite in the G51.21+0.11 complex. To date, the nature of the GeV emission has not been investigated. 
On the basis of our findings a physical relationship between the {\itshape Fermi}  source and the radio emitting components in the field cannot be ruled out. It could represent a new case of a Galactic $\gamma$-ray emitting source related to a complex including the remains of a stellar explosion and star-forming activity in the HII regions.

\section{Concluding remarks}
Low radio frequency observations from the VLSSr at 74~MHz have allowed us to distinguish the nonthermal emission component in a Galactic complex encompassing considerable thermal emission. While earlier work characterised the entire $\sim$30$^{\prime}$ complex as the SNR candidate G51.21+0.11, we have been able to isolate the main region of nonthermal emission to less than 10\% of the complex. This allows us to reclassify the SNR candidate as G51.04+0.07 (G51.04). 
The computed flux density integrated over the whole extent of the emission at 74~MHz is 6.1$\pm$0.8~Jy. The unambiguously nonthermal global spectral index ($\alpha$$\sim$$-$0.52) favours an interpretation in which the detected radiation is produced by a supernova remnant probably belonging to the shell-type morphology class. 
The integrated synchrotron flux density is $\sim$1.51~Jy at the nominal frequency of 1~GHz, making G51.04 among the $\sim$15\% of Galactic SNRs whose 1~GHz integrated flux densities are below 2~Jy. This is unsurprising since observational selection effects undoubtedly hide many weakly emitting SNRs within numerous inner Galactic complexes. In fact, we cannot rule out that future observations will reveal that the larger complex G51.21+0.11 may harbour additional SNRs. In addition, the lack of a turnover in the integrated spectrum of G51.04 below 100 MHz indicates that the newly defined SNR is closer than the HII regions located in the associated complex at about 8~kpc from us, and possibly immediately on their near side. 
Our discovery highlights the crucial importance of low radio frequency observations with sufficient angular resolution and surface brightness sensitivity in addressing the `missing SNR problem' in our Galaxy. Addressing this challenge is key to understanding the star formation history of the Galaxy and bears on central problems including the origin of Galactic cosmic rays and the structure and evolution of the interstellar medium.

We have also discovered an elongated HI cavity-like feature whose location and morphology indicate interaction with G51.04. The physical parameters of this structure were used to constrain the kinematic distance and age of the radio continuum source to be equal to $\sim$7.7~kpc and $\sim$6300~yr, respectively. 
Additionally, we identified molecular material that is matching (in projection) the central portion of the SNR, where an enhancement in the radio continuum emission is observed. Despite the suggested morphological implications, the astrophysical connection of the molecular gas emission with the SNR is not straightforward. 
The correspondence between radio spectral index distribution with the molecular and nonthermal emission at 74~MHz for the brightest central portion of G51.04, suggests the SNR shockwave is encountering molecular material, though not necessarily implying a strong interaction. The situation might be different along the ridge of G51.04, where the flat spectrum component suggests the SNR shock is being decelerated during its passage across the dense interstellar material. We did not find convincing spectral evidence of ionisation processes delineated by 74~MHz free-free absorption in this part of the SNR. However such signatures are subtle and higher resolution, low-frequency observations are probably required to detect them, if present.

\bibliographystyle{aa}
 \bibliography{g51.04}

\begin{acknowledgements}
The authors wish to thank the referee Dr. David J. Helfand for his insightful comments and detailed corrections to the manuscript. 
G. Castelletti is members of the {\itshape Ca\-rre\-ra del Investigador Cient\'{\i}fico} of CONICET, Argentina. L. Supan is a post-doc Fellow of CONICET, Argentina.
This research was partially supported by grants awarded by ANPCYT (PICT~1759/15) and the University of Buenos Aires (UBACyT 20020150100098BA), Argentina.
Basic Research at the Naval Research Laboratory is funded by 6.1 base programs. 
This publication makes use of molecular line data from the Boston University-FCRAO Galactic Ring Survey (GRS) and of data products from the Midcourse Space Experiment. 
\end{acknowledgements}

\end{document}